# Electron spin contrast of Purcell-enhanced nitrogen-vacancy ensembles in nanodiamonds

S. Bogdanov[*,†], M. Y. Shalaginov[*,†], A. Akimov[‡,§,k], A. S. Lagutchev[†], P. Kapitanova[¶], J. Liu[**], D. Woods[*,†], M. Ferrera[††], P. Belov[¶], J. Irudayaraj[**], A. Boltasseva[*,†], and V. M. Shalaev[*,†,‡]

[*]School of Electrical and Computer Engineering, Purdue Quantum Center, Purdue University, West Lafayette, IN 47907, USA

[†]Birck Nanotechnology Center, Purdue University, West Lafayette, IN 47907, USA

[‡]Department of Physics and Astronomy, Texas A&M University, College Station, TX 77843 USA

[§]Russian Quantum Center, ul. Novaya 100, BC "Ural", Skolkovo Innovation Center, Moscow Region, 143025, Russia

[k]Lebedev Physical Institute RAS, Leninskij pr. 53, Moscow, 119991, Russia

[¶]The International Research Centre for Nanophotonics and Metamaterials, ITMO University, Saint Petersburg 197101, Russia

[**]Agricultural and Biological Engineering, Purdue University, West Lafayette, IN 47907, USA

[††]Institute of Photonics and Quantum Sciences, SUPA, Heriot-Watt University, Edinburgh EH14 4AS, United Kingdom

Nitrogen-vacancy centers in diamond allow for coherent spin state manipulation at room temperature, which could bring dramatic advances to nanoscale sensing and quantum information technology. We introduce a novel method for the optical measurement of the spin contrast in dense nitrogen-vacancy (NV) ensembles. This method brings a new insight into the interplay between the spin contrast and fluorescence lifetime. We show that for improving the spin readout sensitivity in NV ensembles, one should aim at modifying the far field radiation pattern rather than enhancing the emission rate.

Nitrogen-vacancy color centers (NV) in diamond are fluorescent lattice defects resulting from a vacancy and an adjacent nitrogen substitution [1,2]. These color centers have proven to be excellent testbeds for novel nanoscale optical devices. Ultrasensitive electromagnetic field [3–8], strain [9,10], pressure [11], and temperature [12,13] sensors as well as integrated quantum information processors [14–16] operating at ambient conditions have been prototyped using NVs. These capabilities are in large part due to the unique properties of the NV's electron spin, which may be optically initialized and manipulated by microwave signals [17,18]. The NV exhibits a spin-dependent fluorescence rate, which can be used for optical spin state readout [19]. The relative difference between the fluorescence rates emitted by the $m_s = 0$ and $m_s = \pm 1$ (where $m_s$ is a spin projection), is commonly called the spin contrast. This spin contrast constitutes the readout signal for spin-based qubits and sensors. Numerous potential applications of NVs such as nanoscale magnetometry or quantum information processing demand the optimization of the spin readout. Such optimization should take into account both the overall photon detection rate and the magnitude of the spin contrast.

The observed fluorescence intensity is typically limited by the inefficiency of photon collection. To combat this inefficiency, various photonic and plasmonic approaches have been tried such as solid immersion lenses [20–22], photonic nanowires [23], cavities [24–28], plasmonic apertures [29], nanoantennas [30,31], waveguides [32–34] and metamaterials [35]. These structures work by modifying the near-field and far-field behavior of the emission thus drastically enhancing the collection efficiency. Additionally, when optically coupled to a photonic resonator and/or a plasmonic structure, the NV center exhibits a reduction of fluorescence lifetime. This reduction is called the Purcell effect and results from a high local photonic density of states (PDOS) [36]. This effect can improve NV's quantum efficiency, leading to even higher photon detection rates. However, despite the vast knowledge accumulated about the NV level dynamics [1], the effect of the fluorescence lifetime on the spin contrast remains unclear. The dependence of spin contrast on fluorescence lifetime has been investigated theoretically using different models [37,38]. Here, we present the first experimental study that quantitatively explores this dependence.

The spin contrast in single NV centers monotonically increases with the optical excitation rate and therefore, it is usually advantageous to operate isolated NV centers in the optical saturation regime. However, for sensing applications such as magnetometry, one often chooses to employ NV ensembles (NVEs) with inter-defect separation distances (IDSD) on the order of 10 nm and smaller [39–42], yielding high levels of fluorescence. Unlike single NV centers, these ensembles must be operated at optical excitation rates well below the saturation level because the spin contrast exhibits an optimum well before the saturation regime is reached. This observation is also confirmed by unpublished measurements conducted in other groups [43]. In this work, we measure the dependence of the spin contrast on the fluorescence lifetime in dense NV ensembles ($IDSD \approx 8$ nm). We also explore the implications of this dependence for the design of NV-based nanophotonic devices. For this study, we introduce a novel technique for spin contrast measurement that is particularly suited for such NVEs.

In our experiment, individual nanodiamonds (76 ± 20 nm in size), each containing an NVE (400 NV centers, on average), were dispersed on a sapphire substrate. In order to create a wide distribution of fluorescence lifetimes, 0.5 mm diameter plasmonic titanium nitride (TiN) [44] islands were formed on the substrate. The NVEs were experiencing different PDOS depending on their location. Higher PDOS at the surface of TiN islands is expected due to confined surface plasmon-polariton (SPP) modes [45]. Figure 1 (a) shows the layout of the sample and probed areas. We chose an area on sapphire and an area on a TiN island, randomly selected approximately 10 nanodiamonds from each area and measured their fluorescence lifetimes and spin contrast values.

To reduce the number of experimental uncertainties affecting the measurement of the spin contrast, we have devised a novel method based on the process of thermal spin relaxation. First, an initializing optical pulse (see Figure 1(c)) projects the spin into the $m_s = 0$ state. After a controlled time delay $\Delta t$, part of the population relaxes back to the $m_s = \pm 1$ states (see Figure 1(b)). Finally, the 'read' pulse is applied, and the fluorescence is collected during the first $t_{\text{det}} = 300$ ns of the read pulse (see Figure 1(c)). The delay $\Delta t$ is varied to produce different spin populations, starting from a predominantly (70 to 90% [46–48]) $m_s = 0$ spin and ending with a thermally relaxed spin (1/3 of the population in the $m_s = 0$ state). As $\Delta t$ surpasses the spin relaxation time $T_1$, the contrast between the relaxed spin and the initialized spin asymptotically reaches a constant value corresponding to a complete thermal spin relaxation. We refer to this limit value as the $T_1$ spin contrast: $C_{T1} = (N_\infty - N_0)/N_0$. Here, $N_0$ and $N_\infty$ are the numbers of detected photons in the cases of initialized spin and a fully thermalized spin, respectively. Typical spin relaxation curves for NVEs on sapphire and on TiN are shown in Figure 2 (a), featuring spin relaxation times in the 100 µs range.

Unlike a conventional spin contrast measurement based on coherent spin population inversion [49], this technique is advantageous for large NVEs. A resonant microwave pulse would only address a group of NV centers having the same projection on the axis of the DC magnetic field. In contrast, thermal relaxation equally affects all the NV centers in the ensemble, leading to 1/3 of the whole spin population residing in the $m_s = 0$ state. $C_{T1}$ measured on an NVE represents 2/3 of the spin contrast $C_\pi$ obtained from Rabi oscillations of a single NV center. The measurement of $C_{T1}$ is not affected by strong spin decoherence rates present in dense ensembles. Finally, it does not require the application of DC and AC magnetic fields and therefore is not affected by their temporal and spatial variations.

The fluorescence lifetime measurements were performed using the time-correlated single-photon counting (TCSPC) technique [50]. The fluorescence decay curves for NVEs on sapphire and on TiN are shown in Figure 2 (b). The fluorescence decay data is fitted by sums of exponential decays for ensembles of two-level systems, assuming that their lifetimes are gamma-distributed [51].

We correlated the fluorescence lifetimes and spin contrasts for a collection of NVEs found on the sapphire and TiN areas (see Figure 3). The range of fluorescence lifetimes for NVEs found on sapphire spans 15 to 24 ns. This spread of lifetimes can be attributed to several effects. For example, variations of local density of states experienced by different NVEs [52,53] due to variations in nanocrystal sizes and shapes as well as varying direct nonradiative decay rates [54] can affect the observed ensemble lifetimes. The TiN film's SPP modes [55] contribute to the local PDOS [51] and increase the radiative rates of the NVEs [35]. Correspondingly, the lifetimes measured on TiN area range from 7.5 to 12.5 ns. We have found that the spin contrast strongly depends on the fluorescence lifetime with values of spin contrast $C_{T1}$ dropping to below 5% for the NVEs with the shortest lifetimes.

We have calibrated the laser power using saturation measurements to ensure that both nanodiamonds on sapphire and TiN experience a similar optical excitation rate $k_{\text{opt}} \approx 1.5 \text{ MHz}$ [51]. For larger pump powers, we found that the contrast $C_{T1}$ drops and almost completely vanishes in strong saturation [51]. The origin of this effect, which is still under investigation, can be attributed to the charge exchange processes involving proximal NV centers and/or nitrogen impurities. Such dynamics may be especially pronounced in dense ensembles like ours, e.g. due to Auger-type effects. This effect makes it impractical to work in the saturation regime and limits the observable spin contrast values.

Before rigorously investigating the observed dependence of spin contrast on fluorescence lifetime, we present a qualitative explanation, based on the NV level structure (Figure 4(a)). For simplicity, in this discussion we assume that the optical excitation rate is much lower than all the level decay rates. Following the absorption of a photon, both the excited state (ES) and the singlet levels relax into the ground state (GS) levels before the next photon is absorbed. The excited states (ES) of the $m_s = 0$ and $m_s = \pm 1$ subsystems (i.e. levels $|0e\rangle$ and $|1e\rangle$ respectively) have equal radiative decay rates ($k_{rad}$) into their respective ground states (GS) $|0g\rangle$ and $|1g\rangle$. However, the fluorescence rate of the $m_s = 0$ subsystem is higher, because the non-radiative decay of $|0e\rangle$ through the singlet state $|s\rangle$ is less probable than that of $|1e\rangle$ ($k_{cross}^{(0)} < k_{cross}^{(1)}$). Under an optical pulse, two NV centers initially prepared in $m_s = 0$ and $m_s = \pm 1$ states will exhibit different levels of fluorescence (see Figure 4(b)), leading to a spin contrast. The decay $|1e\rangle \to |s\rangle \to |0g\rangle$ through the singlet state is a non-radiative process (intersystem crossing), and its rates $k_{cross}$ and $k_s$ are not sensitive to the PDOS. Shortening the direct decay lifetime leads to a smaller relative probability of the non-radiative decay and therefore, a reduced spin contrast. Hence, for the case of a low excitation rate, one indeed expects to measure smaller spin contrasts in a higher PDOS environment, as illustrated by our data.

The theoretical calculation of the observed dependence requires a careful analysis of the transient populations of the NV under an optical pulse. The evolution of NV level populations with time can be derived from the master equation $\dot{\mathbf{m}} = \mathbf{A}\mathbf{m}$ [56] that governs the kinetics of the NV center transitions. In this equation, $\mathbf{m}^T = \begin{bmatrix} \rho_{0g} & \rho_{1g} & \rho_s & \rho_{0e} & \rho_{1e} \end{bmatrix}$ is the unknown vector consisting of level populations. The matrix $\mathbf{A}$ is given by:

$$\mathbf{A} = \begin{bmatrix} -k_{\text{opt}} & 0 & k_s \cos^2 \Phi & k_{\text{rad}} & 0 \\ 0 & -k_{\text{opt}} & k_s \sin^2 \Phi & 0 & k_{\text{rad}} \\ 0 & 0 & -k_s & k_{\text{cross}}^{(0)} & k_{\text{cross}}^{(1)} \\ k_{\text{opt}} & 0 & 0 & -k_{\text{rad}} - k_{\text{cross}}^{(0)} & 0 \\ 0 & k_{\text{opt}} & 0 & 0 & -k_{\text{rad}} - k_{\text{cross}}^{(1)} \end{bmatrix} \quad (1)$$

Here, $k_{\text{rad}} = \tau_{\text{rad}}^{-1}$ is the rate of spin-conserving direct ES decay, $k_{\text{opt}}$ is the optical pumping rate, $k_{\text{cross}}^{(i)}$ are the intersystem crossing rates from $|0e\rangle$ and $|1e\rangle$ to the singlet state, $k_s$ is the deshelving rate of the singlet state and the angle $\Phi$ quantifies the branching ratio of the singlet state decay. We assume that the spin decay is negligible during the optical pulse duration of 15 μs, which is well supported by the spin relaxation curves on Figure 2(a). The number of photons arriving within the detection window $t_{\text{det}}$ is $N = \int_0^{t_{\text{det}}} k_{\text{rad}} (\rho_{0e}(t) + \rho_{1e}(t)) dt$. Many NV centers are present in each NVE and the nanocrystal lattice orientations are random. Consequently, the calculated NVE fluorescence rates are obtained by integrating the fluorescence rates over the NV axis directions and lifetimes. The distribution of NV axis directions is assumed isotropic and the lifetimes are assumed to follow the gamma distribution [51]. The spin-dependent non-radiative intersystem crossing rates from ES to the singlet state are found to be $k_{\text{cross}}^{(0)} = 5$ MHz and $k_{\text{cross}}^{(1)} = 30$ MHz, and the deshelving rate of the singlet state $k_s = 7$ MHz. The branching of singlet decay corresponds to $\Phi = 33°$. These fitted values agree fairly well with values found in other experiments [56,57]. The radiative ES decay rate $k_{\text{rad}}$ depends on the local environment of each NVE and is determined from TCSPC measurements.

At optical excitation rates, exceeding 1.5 MHz, we observe a deviation of the spin contrast from this kinetic model [51]. In all spin contrast measurements from Figure 3 the laser powers were such that this deviation was negligible. Using the above parameters, a reasonably good agreement with the experiment was reached within the entire range of measured lifetimes (7 ns to 24 ns) as seen from Figure 3.

Our data shows that the shortening of fluorescence lifetime in NVEs results in a decrease of the optical spin contrast at low excitation rates. This in turn affects the electron spin readout sensitivity. The single-shot spin readout signal-to-noise ratio (SNR) can be assessed as $\text{SNR} \approx C\sqrt{N_0/(2-C)}$. Here, $C$ is the spin contrast and $N_0$ is the number of photons collected within the detection window for an NV center initialized in the $m_s = 0$ state. Plasmonic or resonant photonic structures, such as nanoantennas and nanocavities, can increase $N_0$ by improving the apparent quantum yield and collection efficiency thanks to a high PDOS in specific modes. Nevertheless, at low pump powers, the rapid drop in contrast for NV centers with lifetimes below 5 ns represents a serious limitation to the SNR (solid line in Figure 5). Thus, the spin readout improvement for NV ensembles operating below optical saturation would be best achieved by methods that avoid significant shortening of the fluorescence lifetime. For example, solid immersion lenses [20], bulls-eye gratings [58], photonic nanowires [23] or bulk diamond waveguides [27,59] lead to high collection efficiency through the modification of the far-field radiation pattern, without creating a high PDOS in any particular mode.

The negative effect of the lifetime shortening on the spin readout SNR in dense NVEs is due to the fact, that these NVEs must be operated at low optical excitation rates. In our model, we can remove this limitation, by considering the dynamics of a single NV, unaffected by the ensemble effects. Then, at saturating optical powers, the spin contrast predicted by our model only depends on the non-radiative transition rates $k_s$ and $k_{\text{cross}}^{(i)}$ and, therefore, should not depend on the fluorescence lifetime. Consequently, at $k_{\text{opt}} = 1500 \text{ MHz}$, the Purcell effect could improve the spin readout SNR significantly (see dashed line in Figure 5), even with perfect photon collection. This implies that Purcell effect-based collection schemes could be efficiently utilized in diamond crystals with low defect concentration. We note however that these results may be affected by the presence of spin non-conserving transitions [37,51].

In summary, we have studied the dependence of the spin contrast in nanodiamond-based NVEs as a function of their fluorescence lifetime. Lifetimes up to 24 ns were observed for NVEs in a dielectric environment and as short as 7 ns for NVEs in a plasmonic environment, with the corresponding spin contrast $C_{T1}$ values ranging from 18% to 4%. We have developed a novel method for measuring the optical spin contrast in NV ensembles, relying on thermal spin relaxation and involving no microwave and static magnetic fields. The experimentally obtained dependence of the spin contrast on the lifetime can be adequately described by using a linear rate equation-based model. Our results can be used to optimize the spin readout sensitivity of NVEs in various photon collection schemes and pave the way for improved sensing schemes utilizing NVEs.

**Acknowledgements**: The authors thank Mikhail D. Lukin, Arthur Safira, Alp Sipahigil, Renate Landig and Joonhee Choi for helpful discussions, Samuel Peana and Clayton Devault for their assistance with manuscript preparation and Dimitrios Peroulis for providing the microwave testing equipment for preliminary tests. This work was partially supported by the AFOSR-MURI grant FA9550−12−1−0024 and NSF grants MRSEC DMR-1120923 and META-PREM DMR-1205457. A.A. acknowledges the Russian Foundation for Basic Research grant # 14–29–07127.

[‡]shalaev@purdue.edu

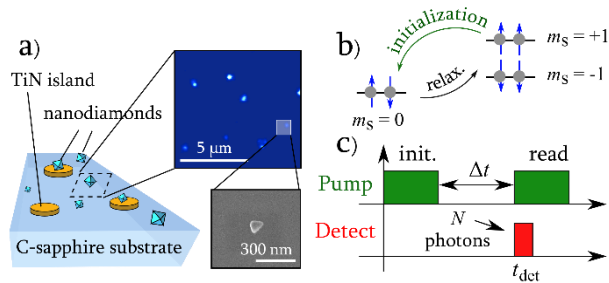

Figure 1 (Color online). (a) Sample layout: islands of plasmonic TiN (200-nm-thick and 500 μm in diameter) and dispersed nanodiamonds with NV center ensembles (NVEs) on a C-sapphire substrate. Blowups show a fluorescence map of NVEs and an SEM image of a typical nanodiamond used in the study. (b) Ground state spin level diagram showing the processes of optical initialization and subsequent thermal relaxation. (c) Spin contrast measurement scheme. The number of photons registered during the detection window of duration $t_{det}$ depends on the degree of spin relaxation occurring during the time $\Delta t$.

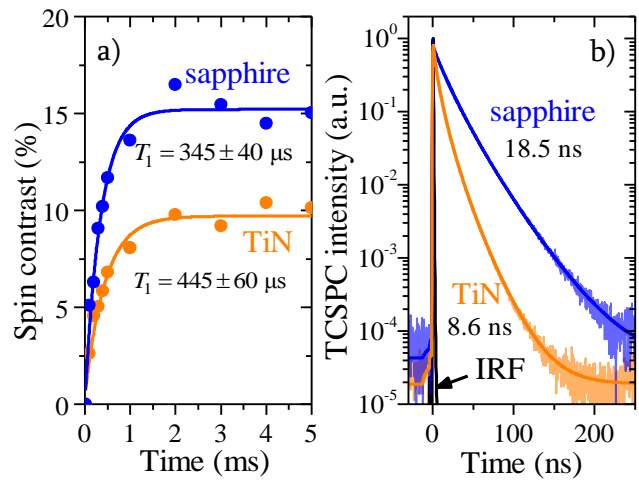

Figure 2 (Color online). (a) Typical spin relaxation and (b) fluorescence decay curves for NVEs found

on sapphire (blue) and TiN (orange). The spin contrast is measured as a normalized difference between the fluorescence signals of partially thermalized spins and optically initialized spins.

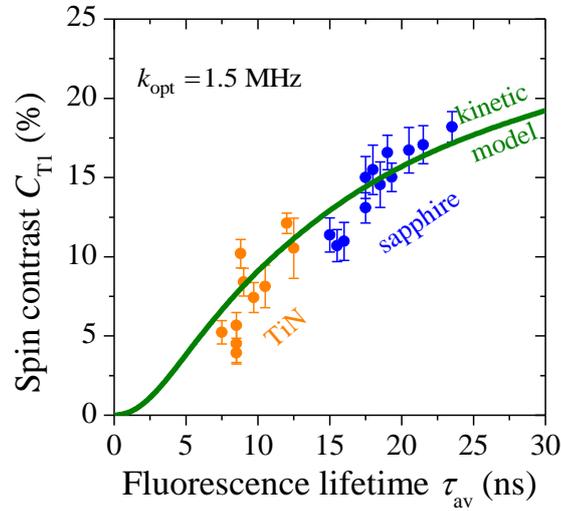

Figure 3. (Color online). Spin decay contrast ($C_{T1}$) values for NVE with different lifetimes measured for an optical excitation rate of approximately 1.5 MHz. The trend agrees well with the results of the simulation based on a kinetic model of the NV.

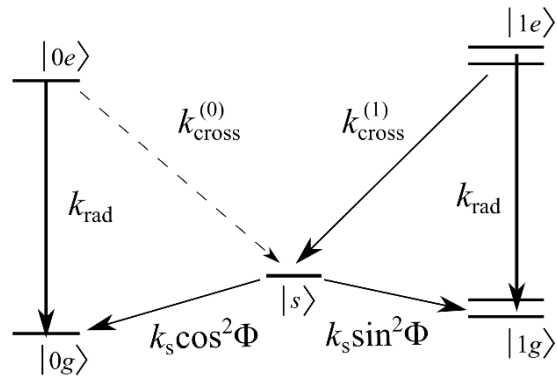

Figure 4. A simplified representation of the NV center's energy levels and transition rates. The two levels on the left (right) are the excited and ground triplet states of the $m_s = 0$ (±1) subsystem. The level $s$ in the middle is the metastable spin singlet level.

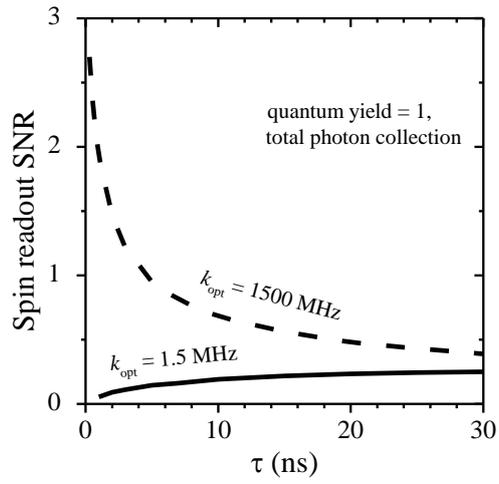

Figure 5. Dashed line: single-shot signal-to-noise ratio of a single NV electron spin readout as function of the total fluorescence lifetime, assuming unity quantum yield and total collection of fluorescence, $k_{opt} = 1500\text{ MHz}$ (deep saturation). Solid line: SNR of the NVE under optical excitation rate of $k_{opt} = 1.5\text{ MHz}$ used in our experiment. The duration of detection window $t_{det}$ is optimized for each datapoint.